\documentclass[cits]{JINST}
\pdfoutput=1

\usepackage[pdftex]{graphicx}

\usepackage[squaren]{SIunits}

\usepackage{amsmath,amssymb}

\graphicspath{{images/}}

\title{Modeling the frequency response of microwave radiometers with QUCS}
\author{Andrea Zonca$^a$\thanks{Corresponding author.}~, Bastien Roucaries$^b$, Brian Williams$^a$, Ishai Rubin$^a$, Ocleto D'Arcangelo$^c$, Peter Meinhold$^a$, Philip Lubin$^a$, Cristian Franceschet$^d$,Stefan Jahn$^e$, Aniello Mennella$^d$, Marco Bersanelli$^d$ \\ 
\llap{$^a$}Department of Physics, University of California, Santa Barbara\\
Santa Barbara, CA 93106, USA\\
\llap{$^b$}
Universit\'e Paris-Est, Laboratoire Central des Ponts et Chauss\'ees\\
75732 Paris, France \\
\llap{$^c$}IFP-CNR,\\
  via Cozzi 53, 20125 Milano\\
\llap{$^d$}Dipartimento di Fisica, Universit\'a degli Studi di Milano\\
Via Celoria 16, 20133 Milano, Italy\\
\llap{$^e$}Infineon Technologies AG,\\
  Am Campeon 1-12, 85579 Neubiberg, Munich, Germany\\
  E-mail: \email{zonca@deepspace.ucsb.edu}}

\abstract{
Characterization of the frequency response of coherent radiometric receivers is a key element in estimating the flux of astrophysical emissions,
since the measured signal depends on the convolution of the source spectral emission with the instrument band shape.

Laboratory Radio Frequency (RF) measurements of the instrument bandpass often require complex test setups and are subject to a number of systematic effects 
driven by thermal issues and impedance matching, particularly if cryogenic operation is involved.

In this paper we present an approach to modeling radiometers bandpasses by integrating simulations and RF measurements of individual components.
This method is based on QUCS (Quasi Universal Circuit Simulator), an open-source circuit simulator, which gives the flexibility of choosing among the available devices, implementing new analytical software models or using measured S-parameters.
Therefore an independent estimate of the instrument bandpass is achieved using standard individual component measurements and validated analytical simulations.

In order to automate the process of preparing input data, running simulations and exporting results 
we developed  the Python package \texttt{python-qucs} and released it under GNU Public License .

We discuss, as working cases, bandpass response modeling of the COFE and {\sc Planck} Low Frequency Instrument (LFI) radiometers and compare results obtained with QUCS and with a commercial circuit simulator software.
The main purpose of bandpass modeling in COFE is to optimize component matching, while in LFI they represent the best estimation of frequency response, since end-to-end measurements were strongly affected by systematic effects.

}


\begin{document}

\section{Introduction}

An accurate characterization of the bandpass response of microwave radiometers is fundamental in the interpretation of a wide range of astrophysical observations at millimeter wavelengths.
In particular, bandpass uncertainties may significantly impact Cosmic Microwave Background (CMB)  experiments, since the combined emission of CMB and the Milky Way galaxy can induce systematic effects both in temperature and polarization measurements.

End-to-end measurements of the bandpasses, however, often require complex test setups that  might be affected by systematic effects
like thermal issues in the cryogenic setup or impedance mismatch in injecting the test signal.

Often, a better insight of the radiometer frequency response can be achieved by building a bandpass model
based on the S-parameters \cite{microwave_pozar} of individual components,  obtained either by measurement or by analytical
simulations.
Such a model can give a robust independent estimate of the integrated radiometer
bandpass,  as it is based on hardware data acquired with much simpler and standard
setups, where systematics are under control.

The comparison of the bandpass model to end-to-end measurement or to sub-assemblies is valuable to identify possible unexpected interactions between the components after integration. Such interactions might then be modelled and included in the bandpass model.

\subsection{Radiometers}

State-of-the-art microwave detectors in the 1-90 GHz frequency range are mainly based on coherent radiometers,
that can achieve a sensitivity below $250$ $ \mu K \sqrt{s}$ (at 30 GHz) thanks to the performance
of cryogenically cooled Low Noise Amplifiers based on High Electron Mobility Transistors \cite{Davis2009}.

One of the most important astrophysical applications of microwave radiometers is in the
observations of the CMB. Since its first discovery in 1964\cite{PenziasWilson}, an uninterrupted chain of radiometric observations have been carried out to measure the CMB spectral shape, anisotropy and polarization properties with increasing precision. The most notable applications were in space missions, starting from COBE-DMR (COsmic Background Explorer-Differential Microwave Radiometer) \cite{Smoot1990} launched in 1989, with 3 channels at 31.5, 53 and 90 GHz, WMAP (Wilkinson Microwave Anisotropy Probe,  launched in 2000) \cite{Jarosik2003} with
5 channels at 23, 33, 41, 61 and 94 GHz, and the {\sc Planck} Low Frequency Instrument \cite{Bersanelli2009}, launched in 2009, with 3 radiometric
channels at 30, 44, 70 GHz.

\begin{figure}[h]
    \centering
    \includegraphics[width=.9\textwidth]{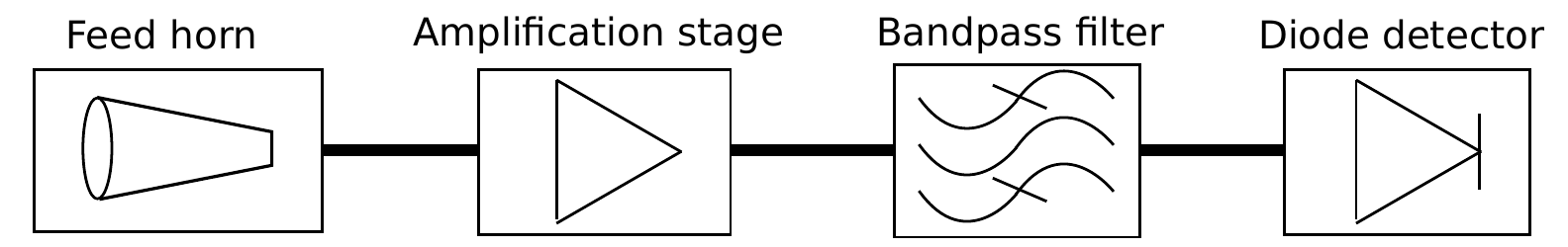}
    \caption{Schematic of a total power radiometer}
    \label{fig:simpleradiometer}
\end{figure}

Radiometer sensitivity scales as:
\begin{equation}
\dfrac{1}{\sqrt{\Delta \nu}},
\end{equation}
where $\Delta \nu$ is the effective bandwidth, defined as:
\begin{equation}
\Delta \nu = \dfrac{(\int g(\nu)d\nu)^2}{\int g(\nu)^2d\nu},
\end{equation}
where $g(\nu)$ is the frequency response. Therefore, in order to optimize the sensitivity, it is favorable a wide and flat band; with current technology it is reasonable to achieve typically an effective bandwidth of about $20\%$ of the detection frequency, e.g. 6 GHz band at 30 GHz.

Bandpasses do not directly  impact CMB temperature anisotropy measurements, because calibration is usually performed on the CMB dipole signal which has the same black body spectrum of the
anisotropies. However, galactic emissions have different spectral shape that will be weighted differently by the in-band instrument response, so that a poorly known bandpass can introduce a systematic error in distinguishing the CMB from the galactic emission. Furthermore, CMB polarization experiments usually require differencing between independent radiometer channels with polarization orientations at $90^{\circ}$ relative to each other as projected on the sky. Thus, a bandpass mismatch between orthogonally oriented radiometers produces a spurious differential signal that mixes with the sky emission, see \cite{2009LFIpolarisationM6}.

The RF power output of the radiometer RF amplification stages is integrated into a DC voltage by the detector diode, so bandpasses can only be measured with
dedicated tests based on a monochromatic signal sweeping through the band. End-to-end frequency
response tests require complex setups due to cryogenics and signal injection into the horn.
On the other hand,  the frequency response of individual radiometer components (orthomode transducers, low noise amplifiers, waveguide components, etc.) can be tested 
with good accuracy and repeatability, thanks to the much simpler and standard setups required. It is therefore
of high interest to develop methods to combine the measured (or simulated) information from the individual components  in a software model of the instrument yielding a synthesised overall bandpass.

Moreover, the new generation of CMB experiments will require focal plane arrays with hundreds or thousands of receivers in order to achieve the required sub-$\mu K$ sensitivity per-pixel. In this scenario, bandpass characterization based on combination of single component data may turn out to be the only viable method to meet the scientific requirements.

In this paper we present the implementation of a bandpass modeling system based on QUCS (Quasi Universal Circuit Simulator)\footnote{\href{http://qucs.sf.net}{http://qucs.sf.net}}, a Free Software circuit simulator.
The use of QUCS ensures the flexibility
of being able to mix, whenever needed, data from laboratory measurements with analytically simulated components.

Running simulations for several receivers with an interactive interface is error prone and heavily time consuming, for this reason we have developed and released under GPL\footnote{\href{http://www.gnu.org/licenses/gpl.html}{http://www.gnu.org/licenses/gpl.html}} \texttt{python-qucs}\footnote{\href{https://github.com/zonca/python-qucs}{https://github.com/zonca/python-qucs}}, a Python package that allows automated QUCS simulation and data exporting.

A previous version of the bandpass model for {\sc Planck}-LFI radiometers, where this work originated, was implemented in the ADS (Advanced Design System) software package by Agilent as described in \cite{Battaglia2009,Zonca2009}.
The use of ADS suffered several limitations: first of all, as any closed source software, it does not allow the customization and implementation of new features, which is very important for non-standard scientific applications, then the high license costs, troublesome for small experiments, and finally not having an easy method for scripting batch simulation runs with general purpose scripting languages like Python or Perl.

In section~\ref{sec:model} we present a general overview on the bandpass model, in section~\ref{sec:wg} we present the implementation of the rectangular waveguide component, then we show two scenarios of application on the COFE radiometers (section~\ref{sec:cofe}) and on the {\sc Planck} radiometers (section~\ref{sec:planck}).

\section{QUCS bandpass model}
\label{sec:model}

In this section we give an overview of the Quite Universal Circuit Simulator (QUCS) bandpass model, starting from an introduction to QUCS itself, focusing on how it is used to characterize bandpasses, followed by details about the rectangular waveguide model we implemented and the supporting Python tools used to automate the simulation and data export process.

\subsection{QUCS}

QUCS, \cite{QUcsschema}, is an open source electronics circuit simulator software released under GPL license.
It gives  the possibility to set up a circuit with a graphical user interface built upon 
QT libraries\footnote{\href{http://trolltech.com/products}{http://trolltech.com/products}} 
and simulate the signal and noise behaviour of the circuit. It is an alternative to well known Berkeley SPICE\footnote{\href{http://bwrc.eecs.berkeley.edu/Classes/IcBook/SPICE/}{http://bwrc.eecs.berkeley.edu/Classes/IcBook/SPICE/}} and Agilent ADS. 

QUCS is coded in C++ and uses extensive class inference in order to facilitate the implementation
of new components.
The main QUCS modules are:
\begin{description}
    \item[\texttt{qucsator}] the command line circuit simulator, which reads a circuit description in a predefined ASCII format, called netlist, and outputs an ASCII format results file.
    \item[GUI] Graphical User Interface, which is completely independent and makes it possible to draw a circuit using a library of devices or file defined components. The GUI automatically builds the netlist, runs \texttt{qucsator}, parses the results file, and allows the user to easily produce tables and plots.
\end{description}

\subsubsection{Solving method for AC simulations}


QUCS simulator can be used for simulating electrical circuits in time or frequency domain, for linear or non linear analysis. The simulation used in this article is based on an AC analysis of the circuit. The AC analysis is done by using a Modified Nodal Analysis (MNA) method, see \cite{MNA}.

MNA is based on solving the circuit equations in the form of a linear system:

\begin{equation}
[A] x = z.
\end{equation}

The $A$ matrix is only dependent on characteristics of the components, the $x$ vector contains the unknown quantities (node voltage and current through
independent voltage sources), and $z$ represents the vector of sources (voltage and current sources). Simulation is done simply by solving
the system.

The main strength of the MNA method is the stamp approach: each component is characterized by four small matrices and one vector
that are directly inserted into the $A$ and $z$ matrices. Thus the matrices are built from small building blocks
without general knowledge of the full circuit topology.

\subsection{Bandpass characterization}

Radiometric devices are receivers capable of measuring the power of tiny microwave signals by providing strong amplification, typical gains of about $10^7$ in terms of power, and
detection through square law detector diodes. A wide variety of receivers have been used in microwave and millimeter wave astrophysics, from simple total power systems, to Dicke-switched receivers, to various correlation schemes \cite{Kraus} . The simplest radiometer, Fig.~\ref{fig:simpleradiometer}, is composed by a feed horn which conveys the signal to an
amplification stage, a bandpass filter which defines the frequency range and a diode which outputs a DC voltage that is proportional to the power of the incident electromagnetic signal and can be easily measured and digitized.

Their bandpass is defined as the normalized gain as a function of frequency, see \cite{Zonca2009}:
\begin{equation}
    V_{out} = G\int_{0}^{\infty} g(\nu) P_{in}(\nu) d\nu + V_{noise},
\end{equation}
where:
\begin{itemize}
    \item $V_{out}$ is the radiometer voltage output
    \item $G$ is the calibration factor in units $[V/W]$
    \item $g(\nu)$ is the normalized bandpass
    \item $P_{in}$ is the radiometric input power
    \item $V_{noise}$ is the voltage due to the radiometer noise
\end{itemize}

In radiometer frequency response modeling we can devise two main cases:
characterization of only the RF section of a radiometer or end-to-end including the diode.

In the first case QUCS just combines the S-parameters of each device taking into account multiple reflection and outputs the S-parameters of the assembly, $S_{21}$ is the bandpass.

In the second case the diode is quadratic so that the transfer function 
is linear between input power and output volts, so that it is necessary to run AC simulations at two different input temperatures and compute the gradient as a function of frequency:
\begin{equation}
    G(\nu) = \dfrac{Vout_2(\nu) - Vout_1(\nu)}{Win_2(\nu) - Win_1(\nu)}.
    \label{eq:deltag}
\end{equation}

Each component can be modeled by measured data or simulated; measurements data consists of a Touchstone\footnote{Touchstone is a standard text format for describing S-parameters and noise properties as a function of frequency} file with the four S-parameters and an optional noise figure, both as a function of frequency.

In case it is necessary to estimate the bandpass response on a larger frequency span than effectively measured, it is easier to extrapolate on the model than
an end-to-end measurement, since the low and high frequency cutoff sections of the response on the measurement are a non-trivial combination of each device frequency response.
An example is given in section~\ref{sec:omt_extra} for the extrapolation of {\sc Planck}-LFI OrthoMode Transducers response.

\subsection{Rectangular waveguide model}
\label{sec:wg}

Since QUCS is Free Software, it is possible to access the source code and implement functionalities which are missing. In order to model LFI rectangular waveguides we have 
implemented a rectangular waveguide model which was then submitted to the QUCS repository and merged into  the following QUCS release.

In this section we describe the implementation of the waveguide device in QUCS.

\subsubsection{Implementation in QUCS}

The rectangular waveguide model implementation is based on the S-parameter components and has been quite straightforward
thanks to the modular conception of QUCS. The code consists of about 600 lines including extensive comments and the GUI.

The waveguide component was validated using Agilent ADS results with a reference waveguide model defined as follows:
     \begin{itemize}
         \item single rectangular waveguide component
         \item WR-22 (same as 44 GHz LFI waveguides)
         \item 800 mm length
          \item material is gold plated stainless steel
         \item simulation temperature is 20 \degree C
     \end{itemize}

\begin{figure}
    \centering
    \includegraphics[width=\textwidth]{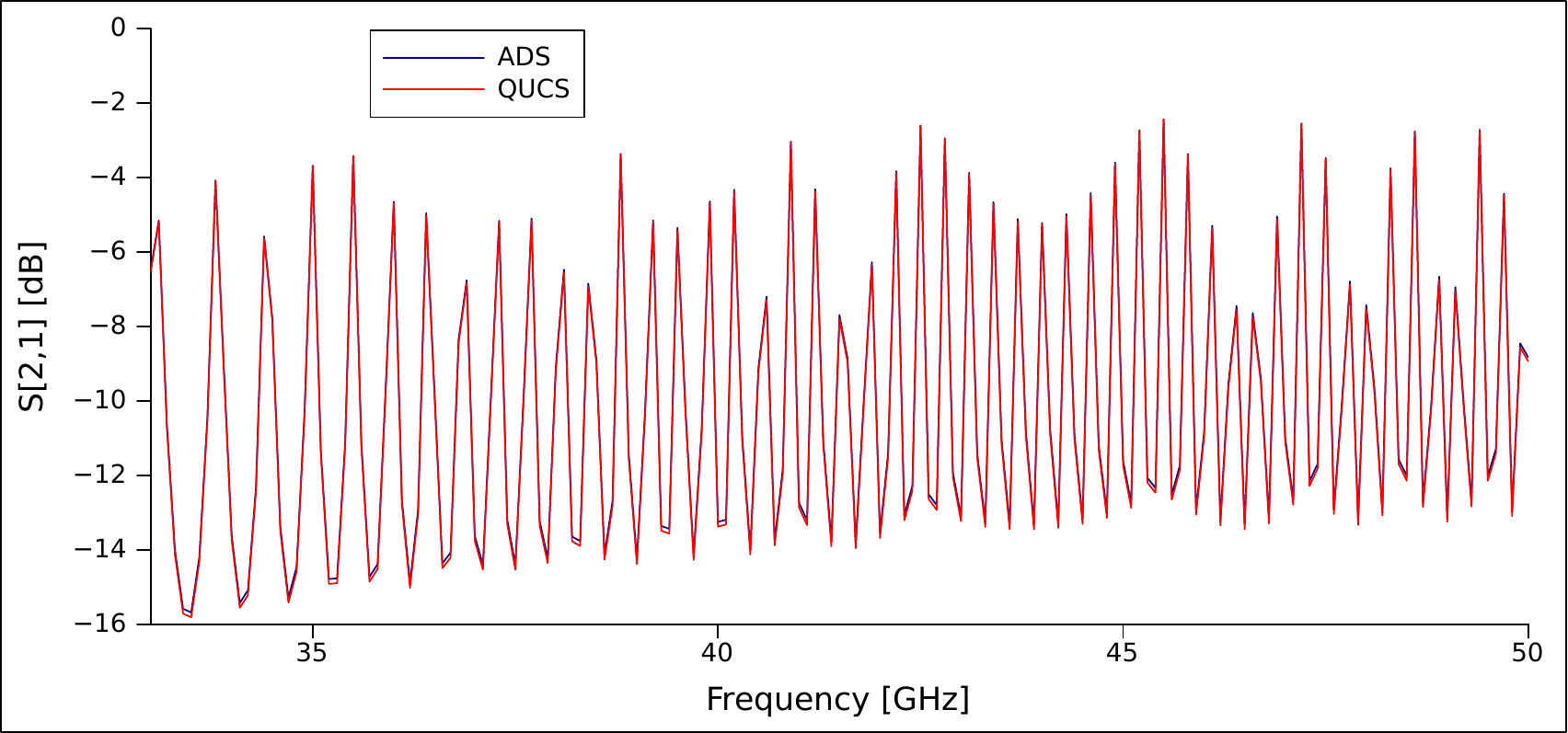}
    \caption{Comparison of S21 [dB] of a gold plated rectangular waveguide as a function of frequency, maximum discrepancy is 0.13 dB}
    \label{fig:adsqucswg}
\end{figure}

Both S parameter and noise simulation voltage results were compared and the matching is at the level of a tenth of a dB, figure~\ref{fig:adsqucswg} shows S21 as a function of input frequency of ADS and QUCS over-plotted.

We also added the embedded computation of the resistivity of gold, stainless steel and aluminum based on device temperature. This feature was implemented using different empirical formulae in different temperature ranges.

\subsection{Batch processing}

In the context of LFI bandpass modeling we developed a set of Python tools for automating the procedure of preparing, running and exporting data from the simulations.

They are based on the software package SciPy\footnote{\href{http://www.scipy.org}{http://www.scipy.org}}, a general purpose scientific and numerical software package for Python, in particular on:
    \begin{itemize}
        \item NumPy: multidimensional array package
        \item Matplotlib: 2D plotting library 
        \item Ipython: interactive Python console
    \end{itemize}
    
The programs are implemented following Object Oriented Programming and are strictly independent, each stage runs separately by reading inputs and writing results in standard ASCII files, the stages are:

\begin{description}
    \item[XLS2S2P:] Format conversion from data stored in an Excel (or Gnumeric) spreadsheet, easier to build and maintain, to Touchstone, used by QUCS.
    \item[\texttt{python-qucs}:] Batch simulation run using \texttt{qucsator} (QUCS GUI is never launched) and exporting of results from QUCS to ASCII format; a simulation of the 44 GHz LFI channels takes about 5 minutes on a common laptop.
    \item[ba\_lib:] Bandpass analysis tool, for batch processing and interactive analysis: 
      \begin{itemize}
        \item gain bandpass from simulations at 2 different temperatures
        \item normalization
        \item plotting
        \item comparison with swept source measurements and ADS simulation
        \item export to text format
      \end{itemize}
\end{description}

These stages were built for {\sc Planck}-LFI simulations but implement several general utilities that can be assembled to build a simulation pipeline for other experiments.

\subsubsection{\texttt{python-qucs}}

\texttt{python-qucs}, the most general tool we developed, was released under GPL on 
Github\footnote{\href{https://github.com/zonca/python-qucs}{https://github.com/zonca/python-qucs}}.
It was designed to easily iterate between testing a circuit on the QUCS GUI and running batch simulations, so that a system can be debugged interactively at several development stages.
The \texttt{simulate} module parses the netlist produced interactively by the QUCS GUI and runs user defined regular expressions to create and save to disk the target set of netlists, then runs \texttt{qucsator} on each of them.
The \texttt{extract} and \texttt{plot} modules allow then to read the custom QUCS file format, manipulate and plot the results.

\section{Application examples: COFE and {\sc Planck}-LFI}
\label{sec:cofe}

In this section we present the application of the QUCS bandpass model to bandpass modeling of two radiometric instruments for CMB measurements: COFE (COsmic Foreground Explorer) and {\sc Planck}-LFI (Low Frequency Instrument).

\subsection{COsmic Foreground Explorer}

COFE, \cite{Leonardi2006}, is a balloon-borne
microwave polarimeter under construction at the University of California, Santa Barbara and designed to measure low-frequency dominant diffuse polarized galactic foregrounds.  The first 24-hour flight, foreseen in summer 2011 from Fort Sumner, New Mexico, will produce temperature and polarization maps at 10 and 15 GHz of 59\% sky with median aggregate sensitivity of $92~\mu K /deg^2$ for temperature and $77 ~\mu K /deg^2$ for polarization.

COFE is comprised of an off axis Gregorian telescope with a reflecting quarter wave plate polarization modulator at the
co-focal point of the two focusing optical elements. Using an external chop to overcome 1/f noise allows COFE to use a total power radiometer.  Each receiver chain has the same design and consists of a cryogenically cooled ($\sim20~K$) InP MMIC LNA \footnote{Indium Phosphide Monolithic Microwave Integrated Circuit Low Noise Amplifier} and provides roughly 30 dB of gain with a noise temperature of 8K. Following the LNA in the cryostat are 2 commercially available GaAs \footnote{Gallium Arsenide} MMICs, a bandpass filter and a square law detector diode. The LNA is connected to the ambient temperature components via low loss coax ($~1 dB$).

\begin{figure}[h]
    \centering
    \includegraphics{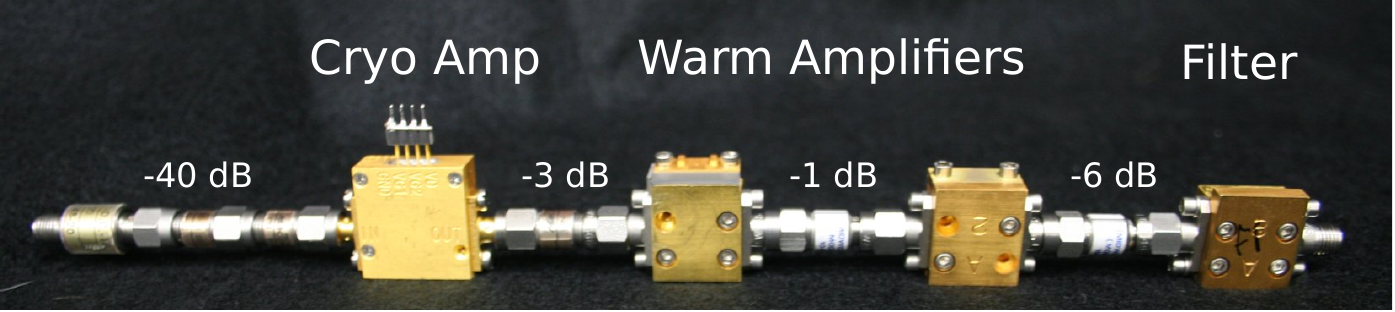}
    \caption{COFE 10 GHz radiometer VNA test setup for bandpass characterization}
    \label{fig:coferad}
\end{figure}
In order to estimate the bandpass response of the RF part of the 10 GHz receiver we performed a dedicated warm Vectorial Network Analyzer (VNA), Rohde\&Schwarz  ZVA40, test on a setup composed of, see figure~\ref{fig:coferad}:
\begin{itemize}
    \item 40 dB attenuation at the input, necessary to reduce the input power generated by the VNA to the expected power level in flight
    \item cryogenic low noise amplifier, $\sim30$ dB gain
    \item 3 dB attenuator, which is the expected loss of the coaxial cable that will connect the cryogenic section to the warm section
    \item first warm amplifier,$\sim20$ dB gain
    \item 1 dB attenuator, which will be present also in flight, useful to provide isolation to the amplifiers
    \item second warm amplifier, $\sim20$ dB gain
    \item 6 dB attenuator designed to provide isolation between the bandpass filter and the amplifying stage
    \item bandpass filter, designed to provide 4 GHz bandwidth and strong out of band rejection
\end{itemize}

All of the components were tested individually and as an assembly with and without the filter on the VNA measuring all S-parameteres.

We built a COFE 10 GHz radiometer model in QUCS using the measured S-parameters of each component and ran simulations in order to have an independent estimate of the bandpass response.

\begin{figure}[h]
    \centering
    \includegraphics[width=.8\textwidth]{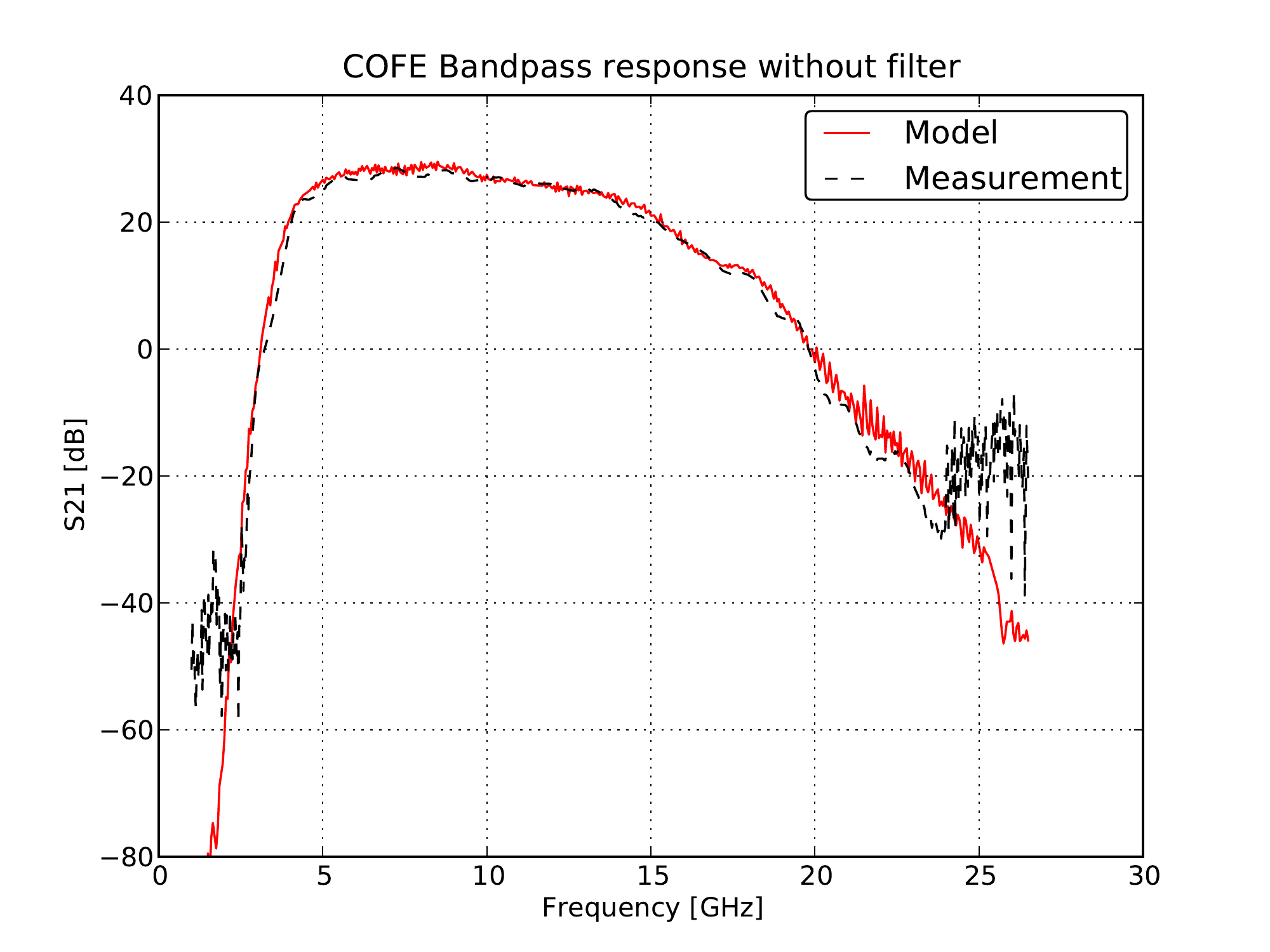}
    \caption{COFE 10 GHz radiometer test and model results without bandpass filter, this includes the $40 dB$ attenuation as part of the setup, the true amplification of the radiometer is almost 70 dB}
    \label{fig:withoutfilter}
\end{figure}

\begin{figure}[h]
    \centering
    \includegraphics[width=.8\textwidth]{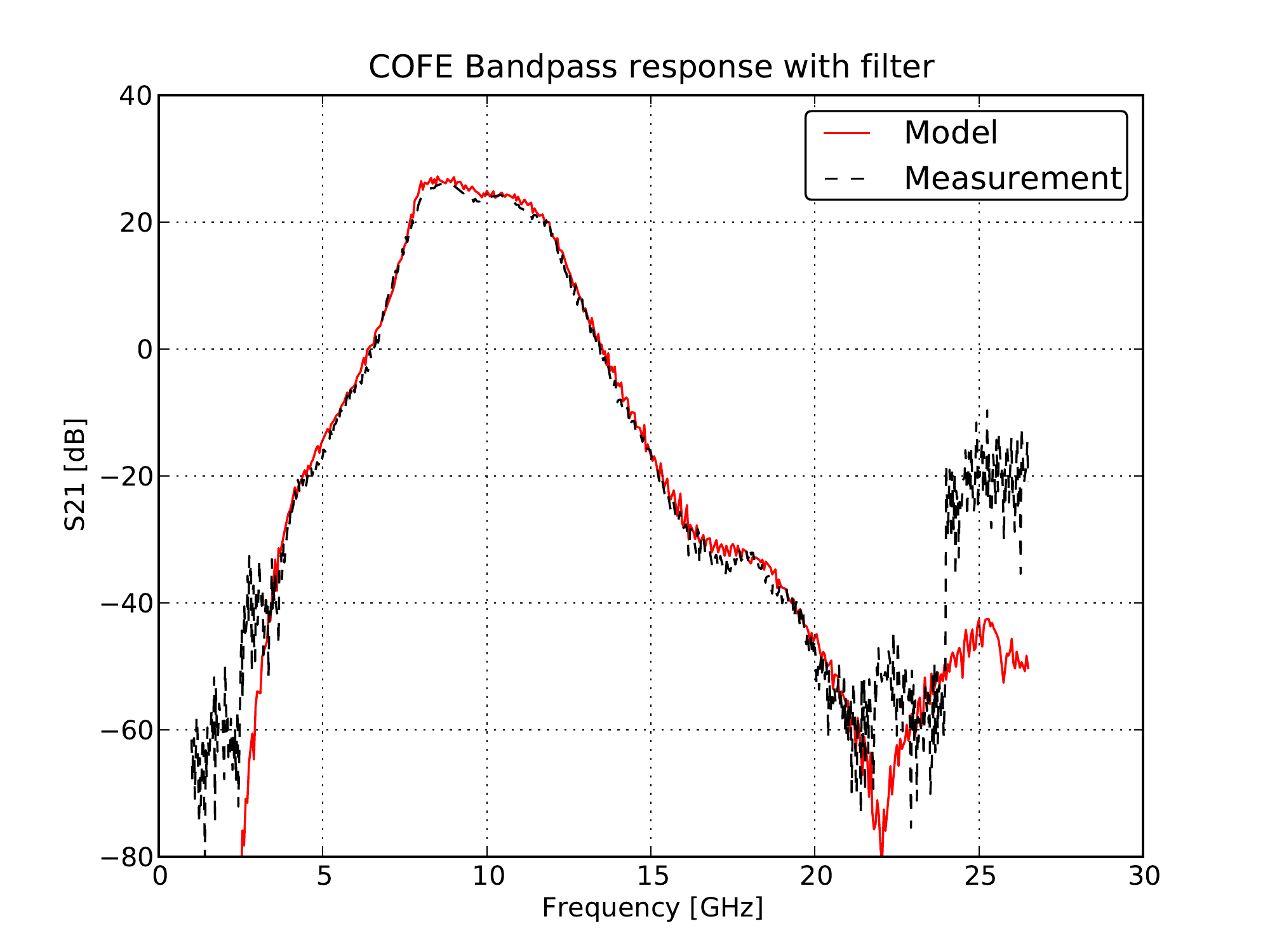}
    \caption{COFE 10 GHz radiometer test and model results with bandpass filter}
    \label{fig:withfilter}
\end{figure}

Figures \ref{fig:withoutfilter} and \ref{fig:withfilter} show the comparison of modeled and measured bandpasses, without and with the filter, respectively. The agreement within the band is better than 1.5 dB for the radiometer with the filter, which is at the same level of the expected accuracy of the measurement. 

The bandpass model proved to give a reliable estimate of the bandpass response of the COFE 10 GHz radiometer. It was also applied to support the filter design and the estimate of the out-of-band rejection.

In the future we plan to use the COFE bandpass model and \texttt{python-qucs} for automating the process of estimating the best combination of the available components. The COFE 10 GHz radiometer has only 3 receivers, but for each of them the cryogenic amplifier can be coupled to any 2 of the 6 available warm amplifiers, their order matters, so the total number of combinations is 90.
We will estimate the bandpass using the model for all the combinations and select just the best in terms of bandwidth for performing hardware testing.

\subsection{{\sc Planck}-LFI}
\label{sec:planck}

As a second example, we consider the 
application of our model to the simulation of the bandpass response of the {\sc Planck}-LFI receivers. We provide an overview of the LFI radiometer scheme, present an example of extrapolation of the response of a device from limited measured data and then compare results of simulations run with QUCS and Agilent ADS \cite{Zonca2009}.

LFI \cite{Mandolesi2010,Bersanelli2009} is a radiometer array mounted in the focal plane of {\sc Planck} satellite telescope, receiving microwave photons at 30, 44 and 70 GHz.

\begin{figure}[h!]
    \includegraphics{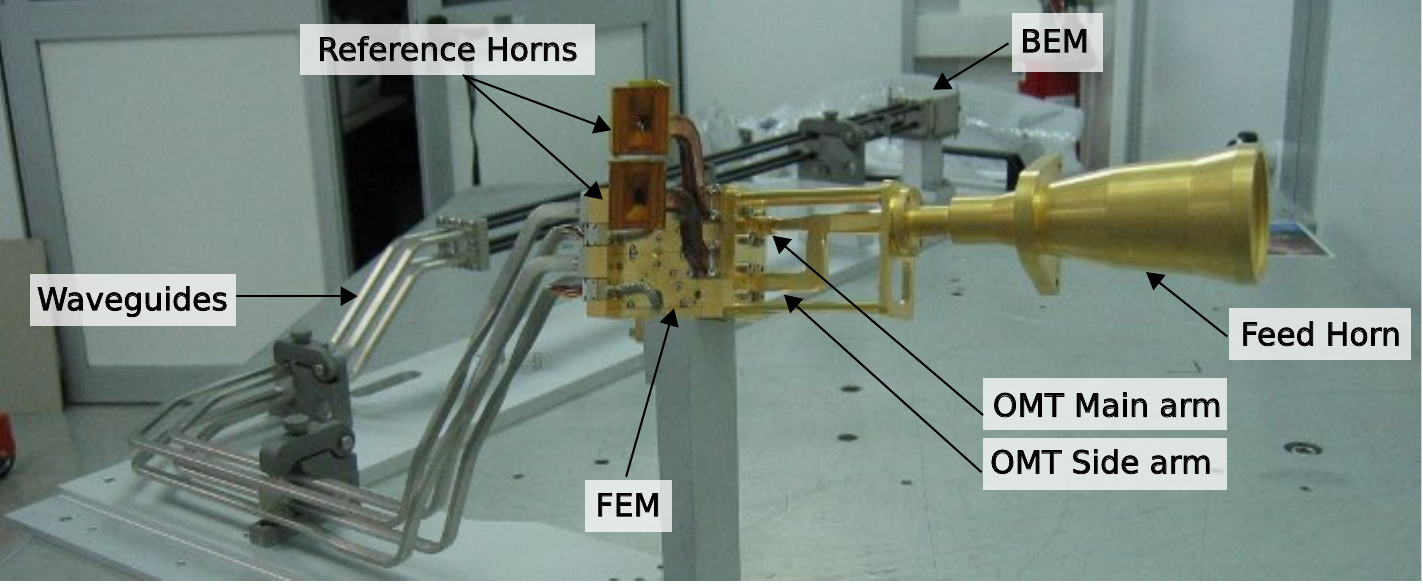}
  \caption{Picture of a 30 GHz Radiometer Chain Assembly (RCA) mounted for testing in Milano}
  \label{fig:rca_photo}
\end{figure}

The complete LFI is an array of 11 Radiometer Chain assemblies (RCA), 6 with a central frequency of 70 GHz, 3 of 44 GHz and 2 of 30 GHz.
Each RCA is composed of (Fig.~\ref{fig:rca_photo}):
\begin{itemize}
\item a Feed Horn looking at the Sky signal
\item  an OrthoMode Transducer (polarisation splitter) 
\item two reference horns optically coupled to two reference loads at about 4K
\item a cold ($\sim$20K) pseudo correlation stage (Front End Module) providing $\sim 35$ dB amplification
\item four waveguides connecting the cold module to the warm backend
\item a second RF amplification stage and bandpass filter
\item a square law detector diode 
\item post-detection DC electronics stage at 300 K (Back End Module)
\end{itemize}

\subsection{Component data extrapolation}
\label{sec:omt_extra}

LFI OrthoMode Transducers (OMT), see \cite{DArcangelo2009}, are passive components based on waveguides that
split the incoming radiation into its orthogonal polarized components. Because the OMT design was exactly the same for 30 and 44 GHz units, 
it is possible to scale at 30 GHz
the measurements performed at 44 GHz, and vice versa. 

In particular we estimated the band between 21.3 GHz and 26.5 Ghz (which was outside the limit for direct measurements) by rescaling 44 GHz measurements which were performed on a larger band, from 33 to 50 GHz.

\begin{figure}[h!]
\includegraphics[width=\textwidth]{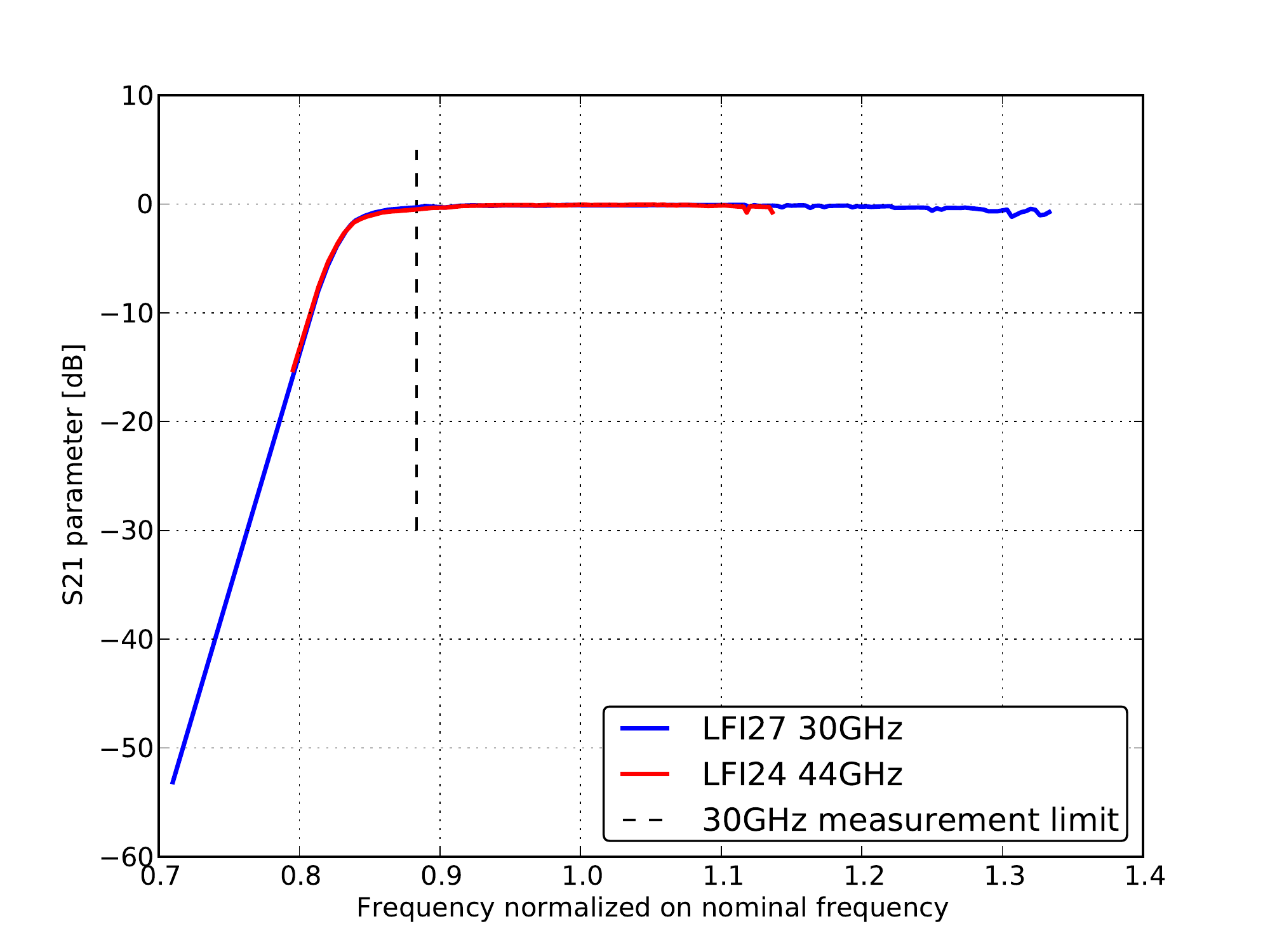}
  \caption{Extrapolation of the S21 30 GHz measurements compared with 44 GHz measurements normalized to their own central frequency. The extrapolation was necessary between 21.3 GHz and 26.5 GHz (black vertical line). At high frequency the cutoff was not characterized because the BEM bandpass filter has strong rejection in that region}
  \label{fig:omt_extra}
\end{figure}

The extrapolation (see figure~\ref{fig:omt_extra}) was done after normalizing the frequency axis on the central frequency, so that all the curves have a central frequency of 1.

\subsection{Model and results}

\begin{figure}[ht]
    \centering
    \includegraphics[width=\textwidth]{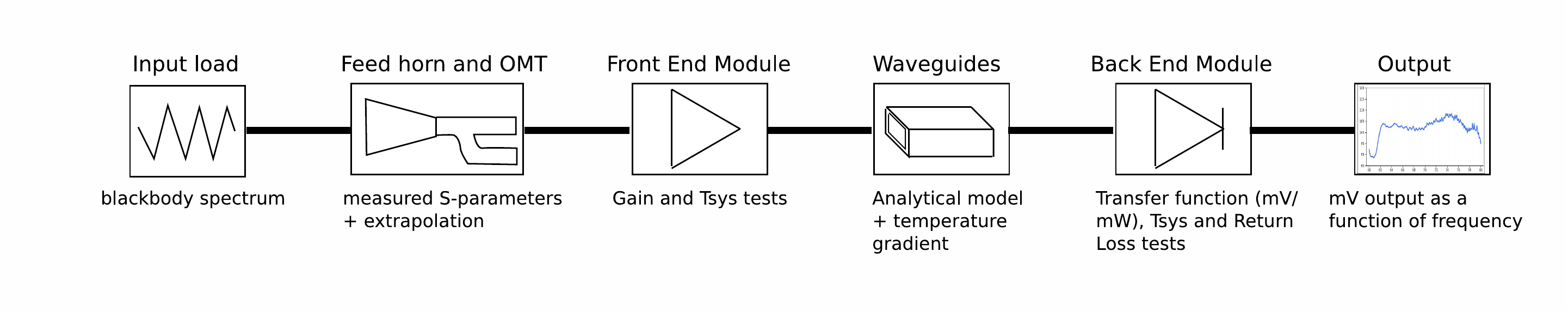}
    \caption{Model schematic: each of the 44 LFI channels is modeled independently as the above model. Each component is represented by its measured behavior as a function of frequency. Waveguides instead are simulated analytically given their dimensions, length and temperature gradient along their length.}
    \label{fig:model_schematic}
\end{figure}

Implementation, excluding the waveguides, faithfully follows the ADS model already described in \cite{Battaglia2009} and \cite{Zonca2009}. For reference we show the model schematic in figure~\ref{fig:model_schematic}.

\begin{figure}
    \centering
    \includegraphics[width=.8\textwidth]{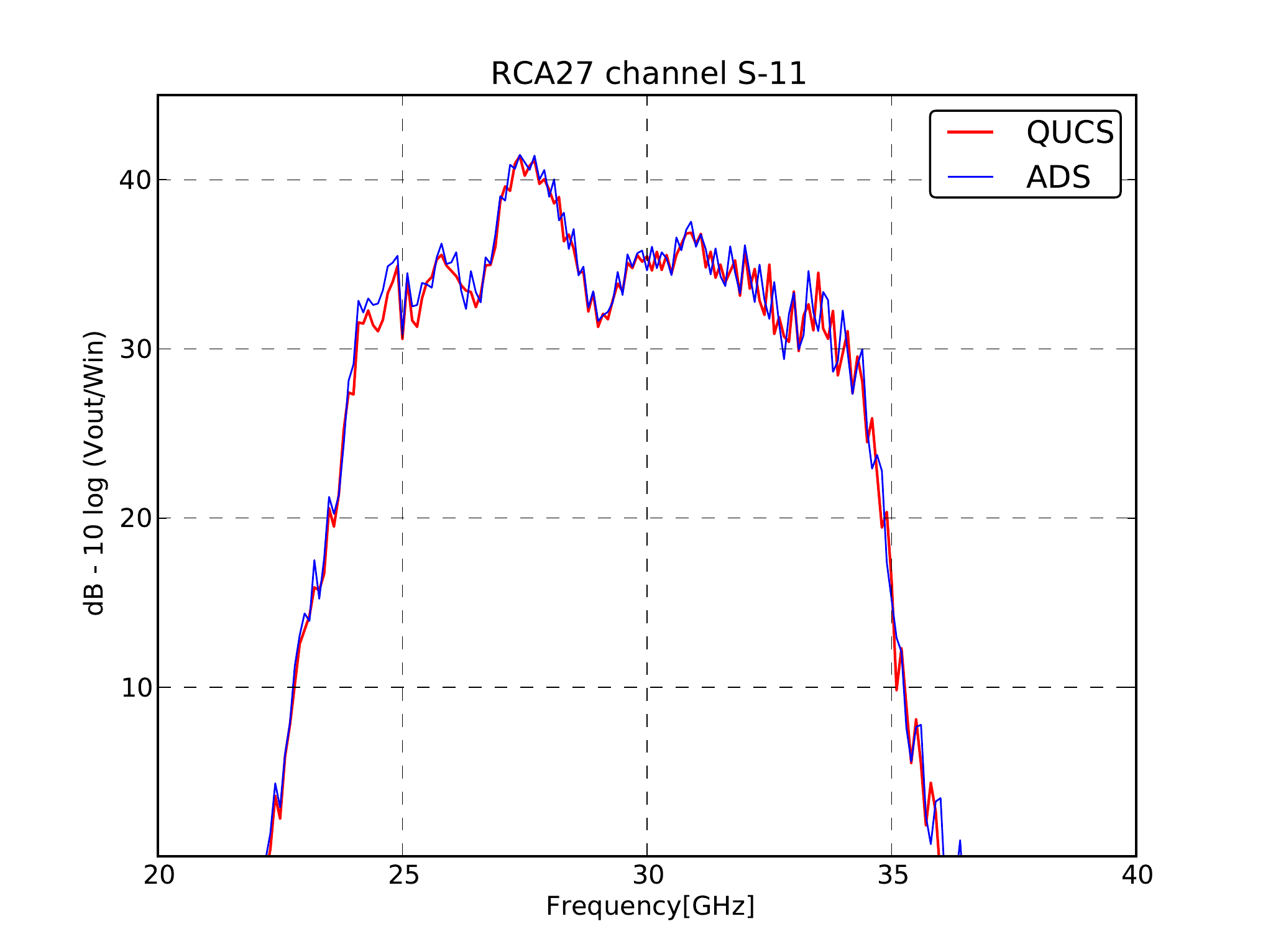}
    \caption{Comparison between ADS and QUCS bandpasses on LFI27S-11 at 30 GHz}
    \label{fig:comp30}
\end{figure}

In figure~\ref{fig:comp30} we show the comparison of estimated bandpasses by ADS and QUCS.
As expected QUCS and ADS results are identical for measured components and differ only for the waveguide simulation, the difference is about $\pm 2$ dB, which is at the same level of the expected bandpasses accuracy.

\section{Conclusion}

In this paper we have presented a method and the software tools capable of estimating radiometer bandpasses from individual component measurements and analytical simulations based on QUCS and Python.
The approach is general and can be used during the design phase of an instrument, gradually replacing analytical simulations, based on requirements, with real data as soon as they become available, in order to continously support frequency response estimation with a software simulation for comparison and validation.

We presented two application scenarios:
in the COFE experiment the model was successfully validated by comparison with VNA measurements and is going to be used to optimize component matching; 
in the {\sc Planck}-LFI mission the model replaced a model based on ADS 
thanks to the possibility of easily batch processing simulations with Python.

\texttt{python-qucs} is a flexible and easy-to-use software tool for making QUCS modeling scale with the number of channels required by the next generation CMB experiments. It provides both powerful batch simulation capabilities for running many channels in different conditions and an interactive debug session for better insight on any issue.

Therefore, QUCS frequency response modeling and \texttt{python-qucs} offer both performance estimation and measurements support in the context of radiometers bandpass estimation, with low cost in terms of budget, since both pieces of  software  are open-source, and in terms of time, since, usually, individual component measurements are already available and the tools strongly reduce the time needed to build and maintain a model.

\acknowledgments
  The work reported in this paper was partly carried out by the LFI instrument team of
  the {\sc Planck} Collaboration.
  Planck is a project of the European Space Agency with instruments
funded by ESA member states, and with special contributions from Denmark
and NASA (USA). The Planck-LFI project is developed by an International
Consortium lead by Italy and involving Canada, Finland, Germany, Norway,
Spain, Switzerland, UK, USA.

  The Italian group was supported in part by ASI through ASI/INAF Agreement I/072/09/0 for the Planck LFI Activity of Phase E2.

  This work was in part supported by NASA under contract number NNG06GC75G, the COFE project.

\bibliographystyle{JHEP}
\bibliography{qucs_article}

\end{document}